\documentclass[prb,onecolumn,superscriptaddress,amsmath,amssymb,floatfix]{revtex4-2}

\usepackage{amssymb,amsmath,epsfig}
\usepackage{subfigure}
\usepackage{color}

 \usepackage[colorlinks=true, pdfstartview=FitV, linkcolor=blue, citecolor=red, urlcolor=magenta, breaklinks=true]{hyperref}
\usepackage{graphicx}
\usepackage{latexsym}
\usepackage{amsmath}
\usepackage{mathrsfs}
\usepackage{amsfonts}
\usepackage{amssymb}
\usepackage{booktabs}
\usepackage{multirow}
\usepackage{verbatim}
\usepackage{orcidlink}
\graphicspath{{Graf/}}
\usepackage{array}

\begin{document}

\title{Classical double copy of black strings in an Anti-de Sitter background}


\author{G. Alencar \orcidlink{0000-0002-3020-4501}}\email{geova@fisica.ufc.br}
\affiliation{Departamento de F\'{\i}sica, Universidade Federal do Ceará\\
Caixa Postal 6030, Campus do Pici, 60455-760, Fortaleza, Ceará, Brazil}

\author{C. R. Muniz \orcidlink{0000-0002-1266-2218}}\email{celio.muniz@uece.br}
\affiliation{Universidade Estadual do Ceará, Faculdade de Educação,\\
Ciências e Letras de Iguatu, 63500-000, Iguatu, Ceará, Brazil}

\author{M. S. Oliveira \orcidlink{0009-0009-2742-2311}}\email{micaeloliveira@fisica.ufc.br}
\affiliation{Departamento de F\'{\i}sica, Universidade Federal do Ceará\\
Caixa Postal 6030, Campus do Pici, 60455-760, Fortaleza, Ceará, Brazil}


\begin{abstract}

We study the classical double copy for static black string solutions in an Anti--de Sitter (AdS) background. By casting the black string metric into Kerr--Schild form over a cylindrical AdS geometry, we construct the corresponding single and zeroth copies. The single copy describes a gauge field satisfying Maxwell-like equations and sourced by an effective line of color charge, while the zeroth copy is given by a scalar field conformally coupled to the AdS background. We also extend the analysis to charged black strings, identifying the associated modifications in the gauge sector. These results show that the classical double copy consistently applies to extended gravitational objects in curved spacetimes.

\end{abstract}
 \maketitle


\section{Introduction}\label{sec_I}

In recent years, considerable efforts have been devoted to investigating the connections between non-Abelian gauge theories and gravity. In this context, the BCJ approach (Bern, Carrasco and Johansson) \cite{Bern:2008qj, Bern:2010ue, Bern:2010yg} provides an elegant formulation of this correspondence within Yang--Mills theory through the color-kinematics duality, allowing gravitational amplitudes to be constructed from gauge-theory amplitudes. At tree level, these relations were demonstrated in \cite{Bern:2010yg, Bjerrum-Bohr:2009ulz} and are consistent with the KLT relations (Kawai, Lewellen and Tye) \cite{Kawai1986}, originally derived in the context of string theory. Taken together, these results indicate that Yang--Mills theory plays a fundamental role in understanding the structure of gravitational amplitudes. In particular, they suggest that, at least perturbatively, gravitational scattering amplitudes admit a double-copy representation in terms of appropriately related gauge-theory amplitudes.

Although originally formulated in the context of scattering amplitudes \cite{Bern:2010ue,Borsten:2020bgv, Kosower:2022yvp, Bern:2019prr}, the scope of the double copy has been successfully extended to classical gravitational solutions. A seminal result in this direction was obtained by Monteiro, O’Connell and White \cite{Monteiro:2014cda}, who showed that the correspondence can be realized exactly for spacetimes admitting a Kerr--Schild (KS) form. Within this framework, the spacetime metric can be written as an exact KS decomposition over a fixed background. The exact linearity of this decomposition, together with the null and geodesic character of the KS vector, constitutes the fundamental property underlying the construction of the classical double copy. This structure naturally induces a hierarchy of associated fields: the so-called single copy, corresponding to a gauge potential, and the zeroth copy, described by a bi-adjoint scalar field. Since its introduction, this classical double copy construction has been extensively explored in a variety of settings, including non-singular black holes \cite{Easson:2020esh}, construction of spacetime metrics through a perturbative expansion \cite{Luna:2016hge}, higher-dimensional black holes \cite{Chawla:2023bsu, Ortaggio:2023rzp}, solutions in three-dimensional space-time \cite{CarrilloGonzalez:2019gof,Gumus:2020hbb,Alkac:2021seh,Gonzalez:2021bes,Alkac:2022tvc}, cosmology \cite{Lee:2022fgr,He:2023iew}, gravitational radiation \cite{Luna:2017dtq}, bigravity \cite{Garcia-Compean:2024zze}, Taub--NUT--type \cite{Luna:2015paa,Bahjat-Abbas:2020cyb,Chawla:2022ogv,Farnsworth:2023mff}, and Kasner \cite{Kent:2025pvu} geometries. Furthermore, the correspondence has been investigated in the context of the fluid-gravity duality \cite{Keeler:2020rcv} and scattering on plane wave backgrounds \cite{Adamo:2017nia}, and also in several other generalizations that have been thoroughly investigated lately \cite{Alkac:2024pfd, Srinivasan:2025hro, Rodriguez:2025hiu}.

Despite the initial success in asymptotically flat backgrounds, generalizing the classical double copy to curved spacetimes presents additional challenges. Recent studies have begun to address the KS formulation in maximally symmetric backgrounds, such as AdS space \cite{Didenko:2008va, Malek:2010mh,Carrillo-Gonzalez:2017iyj, Liang:2023zxo}, which provides a natural setting for these extensions, showing that although the linear structure of the ansatz is preserved, the background curvature plays a crucial role in defining the equations satisfied by the single and zeroth copy fields.

In this work, we investigate the classical double copy applied to solutions with cylindrical symmetry in an AdS background, specifically black strings. Black strings are topological generalizations of black holes with horizon topology $\mathbb{R} \times S^1$, and have been the subject of intense study since the fundamental works of Lemos and others \cite{Lemos:1994xp,Lemos:1995cm}. Furthermore, the treatment of string-like extended objects within the classical double copy formalism has been previously explored in \cite{Goldberger:2019xef}. The goal here is to verify the consistency of the generalized KS construction for these cylindrical geometries with AdS backgrounds and determine the nature of the effective sources in the gauge and scalar sectors.

We first analyze the neutral and static black string, demonstrating that the resulting gauge potential (single copy) corresponds to an electrostatic field generated by an infinite line of color charge, consistent with Maxwell's equations in a cylindrical AdS background. Next, we extend the analysis to the case of the charged black string, where the energy-momentum tensor of the gravitational electromagnetic field induces an effective distributed current in the gauge theory. Furthermore, we discuss the zeroth copy, showing that the scalar profile associated with the KS perturbation satisfies the equation of a scalar field conformally coupled to the AdS geometry.

The paper is organized as follows. In Sec.~\ref{sec_II}, we briefly review the static black string solution in an AdS background. In Sec.~\ref{sec_III}, we recast this geometry into KS form and develop its classical double copy, explicitly constructing the corresponding single and zeroth copies for the neutral configuration and subsequently extending the analysis to the charged black string. Finally, our conclusions are presented in Sec.~\ref{sec_IV}.

\section{Standard black strings} \label{sec_II}

The general form of the metric describing a static black string in a $3+1$-dimensional spacetime, with a negative cosmological constant~\cite{Lemos:1994xp, Cardoso:2001bb}, is given as follows
\begin{equation}\label{eq_metric_bs}
ds^2 = -F(r) \, dt^2 + \frac{dr^2}{F(r)} + r^2 \, d\phi^2 + \alpha^2 r^2 \, dz^2.
\end{equation}
For such a spacetime, the metric~\eqref{eq_metric_bs} possesses cylindrical symmetry, where $0 \leq r < \infty$, $0 \leq \phi < 2\pi$, and $-\infty < z < \infty$ define,
respectively, the radial, angular, and axial coordinates. Furthermore,
$\alpha^2 \equiv -\Lambda/3 > 0$, where $\Lambda$ is the cosmological constant, which appears directly in the action that characterizes the neutral and static black string,
 \begin{equation}\label{eqa}
     S=\frac{1}{2\kappa^2}\int d^4x \sqrt{-g}\, (R-2 \Lambda),
 \end{equation}
where $\kappa^2=8\pi G$ and $R$ is the Ricci scalar, which results in the complete form of Einstein's field equations of general relativity, expressed in the form
\begin{equation}
    G^\mu_\nu + \delta^\mu_\nu \Lambda=\kappa^2 T^\mu_\nu.
\end{equation}
For our purposes, it is sufficient to use the $G_{0}^{0}$ component of the Einstein tensor,
in order to obtain
\begin{equation}\label{eq_temporal}
    G^0_0 = \frac{r F'(r) +F(r)}{r^2}= \frac{1}{r^2}\frac{d}{dr}[r\,F(r)],
\end{equation}
with $F'(r)=dF(r)/dr$, assuming $T^{\mu}_{\nu}=0$ for $r \neq 0$, we obtain
\begin{equation}
    G^0_0 + \delta^{0}_{0} \Lambda = \frac{1}{r^2}\frac{d}{dr}[r\,F(r)]-3\alpha^2=0.
\end{equation}
Solving the differential equation above, we obtain the solution for the function $F(r)$ of the black string spacetime, expressed as follows
\begin{equation} \label{eqF}
    F(r)= \alpha^2 r^2 - \frac{\lambda}{\alpha r}.
\end{equation}
This equation corresponds to the static and neutral black string solution, where $\lambda=4GM$ is associated with the mass per unit length of the black string, and the event horizon is obtained by setting $F(r)=0$, namely,
\begin{equation}
    r \equiv r_{h} = \frac{\lambda^{1/3}}{\alpha}.
\end{equation}

For the charged black string case \cite{Lemos:1995cm}, the gravitational field is generated not only by the cosmological constant but also by an electromagnetic field. The corresponding dynamics is described by the Einstein--Maxwell action with a negative cosmological constant,
\begin{equation}
S=\frac{1}{2\kappa^2}\int d^4x\,\sqrt{-g}
\left(
R-2\Lambda-\frac{1}{4}F_{\mu\nu}F^{\mu\nu}
\right),
\end{equation}
where
\begin{equation}
F_{\mu\nu}
=
\partial_\mu A_\nu-\partial_\nu A_\mu,
\end{equation}
is the electromagnetic field-strength tensor. Varying the action with respect to the metric yields the Einstein field equations,
\begin{equation}
G_{\mu\nu}
+
\Lambda g_{\mu\nu}
=
\kappa^2 T_{\mu\nu}^{\rm EM},
\end{equation}
with
\begin{equation}
T_{\mu\nu}^{\rm EM}
=
F_{\mu\alpha}F_{\nu}^{\ \alpha}
-\frac{1}{4}g_{\mu\nu}
F_{\alpha\beta}F^{\alpha\beta},
\end{equation}
while variation with respect to the electromagnetic potential gives Maxwell's equations,
\begin{equation}
\nabla_\mu F^{\mu\nu}=0.
\end{equation}
For the charged black string solution of Ref.~\cite{Lemos:1995cm}, the electromagnetic four-potential can be written as
\begin{equation}
A_\mu=-h(r)\,\delta_\mu^{0},
\end{equation}
where $h(r)$ is an arbitrary function of the radial coordinate $r$,
\begin{equation}
h(r)=\frac{2Q}{\alpha r}+\mathrm{const.},
\end{equation}
and $Q$ denotes the electric charge per unit length of the black string. Consequently, the only nonvanishing component of the electromagnetic field-strength tensor is
\begin{equation}
F_{tr}
=
-\partial_r A_t(r)
=
-\frac{2Q}{\alpha r^2},
\end{equation}
up to sign conventions for the electromagnetic potential. Therefore, the electromagnetic field acts as the source responsible for the charge contribution appearing in the metric function. Solving the coupled Einstein--Maxwell system yields the charged black string metric function,
\begin{equation}\label{eqF2}
F(r)
=
\alpha^2r^2
-\frac{\lambda}{\alpha r}
+\frac{\beta^2}{\alpha^2r^2},
\end{equation}
where $\lambda$ retains the same interpretation introduced in Eq.~\eqref{eqF} for the neutral solution, while the charge parameter satisfies the relation $\beta^2=4GQ^2$, with $Q$ denoting the electric charge per unit length of the black string. This identification follows directly from the constants introduced in the original solution of Ref.~\cite{Lemos:1995cm}. Therefore, unlike the neutral solution, the charged configuration is not a vacuum solution, since the electromagnetic field contributes nontrivially to the energy--momentum tensor. Furthermore, for distant observers, the solution extends uniformly along the $z$ direction, resulting in a horizon topology of $\mathbb{R}\times S^1$. As in the neutral case, the spacetime possesses a singularity at $r=0$ and is asymptotically AdS in the limit $r\rightarrow\infty$.

\section{The Classical Double Copy}
\label{sec_III}

The correspondence known in the literature as the classical double copy establishes a mapping between classical gravitational solutions and configurations in gauge theories.  A systematic analysis of the classical double copy for statically symmetric KS metrics was developed in \cite{Ridgway:2015fdl}. In addition, several other examples and applications of double copy relations have been obtained, both for flat spacetimes \cite{Monteiro:2014cda,Easson:2020esh,Bah:2019sda} and for extensions to curved backgrounds \cite{Bahjat-Abbas:2017htu, Prabhu:2020avf} and also in dual-field and supergravity theories \cite{Lee:2018gxc,Lescano:2020nve}. Beyond the formulation based on the Kerr--Schild ansatz, which has been extensively applied to Schwarzschild, Reissner--Nordström, Kerr, Kerr--Newman  and Banados--Teitelboim--Zanelli (BTZ) black holes, including extensions to (A)dS backgrounds \cite{Carrillo-Gonzalez:2017iyj,Bah:2019sda,Alkac:2021bav}, alternative approaches to the classical double copy have also been proposed.  In particular, the so-called Weyl double copy \cite{Easson:2021asd, Godazgar:2020zbv, Godazgar:2021iae, Chacon:2021wbr,White:2020sfn, Han:2022mze, Han:2022ubu,Luna:2022dxo, Easson:2022zoh, Alkac:2023glx, Armstrong-Williams:2024bog, Zhao:2024ljb} establishes a direct correspondence involving the gravitational Weyl tensor and has proven particularly successful for type-D gravitational solutions \cite{Luna:2018dpt,Keeler:2020rcv}.

In our analysis, we employ the generalized KS form on an AdS background \cite{Taub1981, Xanthopoulos1983, Malek:2010mh, Ortaggio:2012jd}, which can be written as
\begin{equation}
g_{\mu\nu}=\bar g_{\mu\nu}+2\,\mathcal{H}(r)\,k_\mu k_\nu,
\label{eq_metric_KS}
\end{equation}
where $k_\mu$ is a null vector with respect to both the background (A)dS metric $\bar g_{\mu\nu}$ and the full metric $g_{\mu\nu}$, while $\mathcal H(r)$ encodes the KS sector of the geometry.

Within the classical double copy formalism, starting from a gravitational solution written in the form \eqref{eq_metric_KS}, we define the associated single copy as a gauge potential obtained by replacing one of the null vectors with a constant vector in color space, according to the gravity--gauge correspondence. The associated gauge potential is given by
\begin{equation}\label{regraSC}
A_\mu^{\,a}=
c^{a}\,2\mathcal{H}(r)\,k_\mu .
\end{equation}
where $c^a$ is a constant vector in color space. The zeroth copy, in turn, is simply described by the scalar profile $\mathcal{H}(r)$. These constructions establish a gravity--gauge--scalar correspondence in which the function  $\mathcal{H}(r)$  plays a central role, simultaneously encoding the gravitational perturbation, the associated gauge potential, and the scalar field appearing in the zeroth copy.

\subsection{Neutral black string in Kerr--Schild form}

We then apply the ansatz \eqref{eq_metric_KS} to the neutral and static black string, in order to identify the scalar profile $\mathcal{H}(r)$ and the null vector $k_\mu$ on the cylindrical AdS background.  Our goal is to cast the black string metric, Eq.~\eqref{eq_metric_bs}, into the KS form, and for this first case, we're dealing with a neutral black string. To achieve this, we introduce Eddington--Finkelstein (EF) coordinates, following a procedure analogous to the Schwarzschild metric for black holes in flat spacetime \cite{Visser:2007fj}.  We consider the following ingoing coordinate transformation, 
\begin{equation}
    d\tau = dt + \frac{dr}{F(r)}.
\end{equation}
Substituting this transformation into Eq.~\eqref{eq_metric_bs}, the black string metric takes the form
\begin{equation}\label{eq-metric-ds}
ds^2 =
-F(r)\,d\tau^{2}
+
2\,d\tau\,dr
+
r^{2}d\phi^{2}
+
\alpha^{2}r^{2}dz^{2}.
\end{equation}
This is the complete form of the metric in EF coordinates. To write it in KS form, we begin by separating the metric function $F(r)$ given in Eq.~\eqref{eqF} into its AdS background contribution and the correction associated with the mass parameter. We therefore write
\begin{equation}
F(r)=F_{0}(r)-\frac{\lambda}{\alpha r},
\end{equation}
with
\begin{equation}
F_{0}(r)=\alpha^{2}r^{2}.
\end{equation}
Thus, Eq.~\eqref{eq-metric-ds} becomes
\begin{equation}\label{eq-metrc-full}
ds^2 =
-F_0(r)\,d\tau^{2}
+
2\,d\tau\,dr
+
r^{2}d\phi^{2}
+
\alpha^{2}r^{2}dz^{2}
+\frac{\lambda}{\alpha r}(d\tau)^2.
\end{equation}
From Eq.~\eqref{eq-metrc-full}, we immediately identify the cylindrical AdS background metric in EF coordinates as
\begin{equation}\label{eq-metric-back}
d\bar{s}^{2}
=
-\alpha^{2}r^{2}d\tau^{2}
+
2\,d\tau\,dr
+
r^{2}d\phi^{2}
+
\alpha^{2}r^{2}dz^{2}.
\end{equation}
We can then write
\begin{equation}
ds^2=d\bar{s}^{2}+\frac{\lambda}{\alpha r}(d\tau)^2.
\end{equation}
From the general form of the KS decomposition given in Eq.~\eqref{eq_metric_KS}, we identify the KS vector as
\begin{equation}\label{eq-vetornull}
k_\mu dx^\mu=d\tau.
\end{equation}
which gives
\[
k_\mu k_\nu dx^\mu dx^\nu=(d\tau)^2,
\]
that is,
\begin{equation}
ds^2=d\bar{s}^{2}+\frac{\lambda}{\alpha r}k_\mu k_\nu dx^\mu dx^\nu.
\end{equation}
Comparing this expression with Eq.~\eqref{eq_metric_KS}, we finally obtain the KS scalar function,
\begin{equation}\label{eqfs}
\mathcal{H}(r)
=
\frac{\lambda}{2\alpha r}.
\end{equation}

To ensure that this decomposition is consistent with the standard KS construction, it remains to verify explicitly that the vector $k_\mu$ satisfies the fundamental properties of the ansatz, namely that it is null with respect to the background metric and geodesic. To this end, we use the background metric given by Eq.~\eqref{eq-metric-back}, and from Eq.~\eqref{eq-vetornull} it immediately follows that
\begin{equation}\label{eq-kmu}
k_\mu=(1,0,0,0).
\end{equation}
Raising the index with the inverse background metric, we obtain
\begin{equation}
k^\mu=\bar g^{\mu\nu}k_\nu=(0,1,0,0),
\end{equation}
and, consequently,
\begin{equation}
\bar g^{\mu\nu}k_\mu k_\nu
=
k^\mu k_\mu
=
0,
\end{equation}
confirming that $k_\mu$ is null with respect to the background metric.

Since the components of the vector $k^\mu$ are constant in the adopted coordinates, we have $\partial_\nu k^\mu=0$, so that
\begin{equation}
k^\nu\partial_\nu k^\mu=0.
\end{equation}
Furthermore, since $k^r=1$ is the only nonvanishing component, the covariant derivative along the congruence reduces to
\begin{align}
k^\nu\bar{\nabla}_\nu k^\mu
&=
k^\nu\partial_\nu k^\mu
+
\bar{\Gamma}^{\mu}_{\nu\rho}k^\nu k^\rho \\
&=
\bar{\Gamma}^{\mu}_{\nu\rho}k^\nu k^\rho \\
&=
\bar{\Gamma}^{\mu}_{rr}.
\end{align}
For the background geometry defined by Eq.~\eqref{eq-metric-back}, a direct evaluation of the Christoffel symbols yields
\begin{equation}
\bar{\Gamma}^{\mu}_{rr}=0,
\end{equation}
so that
\begin{equation}
k^\nu\bar{\nabla}_\nu k^\mu=0.
\end{equation}
Therefore, $k^\mu$ defines an affinely parametrized null geodesic congruence with respect to the AdS background metric, satisfying the requirements of the standard KS construction.

Having explicitly identified the KS scalar function $\mathcal H(r)$ and verified that the vector $k_\mu$ satisfies the fundamental properties of the KS ansatz, we are now in a position to construct the corresponding single copy of the solution.

\subsection{The single copy}

We have seen that, in ingoing Eddington--Finkelstein coordinates $(\tau,r,\phi,z)$, the null KS vector takes the simple form given in Eq.~\eqref{eq-vetornull}. By directly applying the standard classical double copy prescription given in Eq.~\eqref{regraSC}, we obtain the gauge potential
\begin{equation}
A_\tau^{a}=c^{a}\frac{\lambda}{\alpha r},
\qquad
A_r^{a}=A_\phi^{a}=A_z^{a}=0.
\label{Acampo}
\end{equation}
The corresponding field-strength tensor describes an electrostatic field with cylindrical symmetry, consistent with an infinitely long linear distribution of color charge along the axis of the black string. The resulting field strength has only one nonvanishing component
\begin{equation}
F_{\tau r}^{a}
=
-\partial_r A_\tau^{a}
=
c^{a}\frac{\lambda}{\alpha r^2}.
\label{Fcampo}
\end{equation}

To verify that the potential \eqref{Acampo} satisfies Maxwell's equations on the cylindrical AdS background, we first consider the source-free sector for $r\neq0$, evaluating
\begin{equation}
\nabla_\mu F^{\mu\nu \, a}=0,
\end{equation}
where, for the AdS background metric in EF coordinates, we have
\begin{equation}
\sqrt{-\bar g}=\alpha r^2,
\qquad
\bar g^{\tau r}=1,
\qquad
\bar g^{rr}=\alpha^2r^2,
\qquad
\bar g^{\tau\tau}=0.
\end{equation}
It then follows that
\begin{equation}
\nabla_\mu F^{\mu\tau a}
=
\frac{1}{\alpha r^2}
\partial_r
\left(
\alpha r^2F^{r\tau a}
\right)=0,
\end{equation}
while the remaining components vanish identically. Therefore, for $r\neq0$, the single copy corresponds to a source-free solution of Maxwell's equations on the AdS background. This configuration describes a purely radial electric field in the transverse plane, uniform along the axial $z$ direction. As in the Schwarzschild case, the associated color charge arises only through a distributional source localized at the origin \cite{Monteiro:2014cda}.

Regarding the interpretation of the source, up to this point we have considered only the region $r\neq0$, which is sufficient to determine the functional form of the metric and the scalar function $\mathcal{H}(r)$. However, as in the Schwarzschild black hole case, the complete solution must be interpreted as being generated by a localized source associated with the mass of the black string. For the spherically symmetric Schwarzschild solution, the gravitational source is described by the energy--momentum tensor of a point source \cite{Monteiro:2014cda,Carrillo-Gonzalez:2017iyj},
\begin{equation}
T^{\mu\nu}
=
M\,u^\mu u^\nu\,\delta^{(3)}(\mathbf{x}),
\end{equation}
which corresponds to a mass concentrated at the origin of three-dimensional space. In the case of the static neutral black string, the situation is analogous, except that cylindrical symmetry implies that the source is not point-like but uniformly distributed along the axial direction $z$. The most general mass density compatible with cylindrical symmetry and with a source localized on the axis can be written as
\begin{equation}
\rho(\mathbf{x}_\perp)
=
\lambda\,\delta^{(2)}(\mathbf{x}_\perp),
\qquad
\int d^2x_\perp\,\rho(\mathbf{x}_\perp)=\lambda,
\end{equation}
where $\mathbf{x}_\perp$ denotes the coordinates transverse to the string. For a static source, the energy--momentum tensor takes the form
\begin{equation}
T^\mu{}_\nu
=
\rho(\mathbf{x}_\perp)\,
u^\mu u_\nu,
\end{equation}
so that
\begin{equation}
T^\mu{}_\nu
=
\lambda\,
u^\mu u_\nu\,
\delta^{(2)}(\mathbf{x}_\perp),
\label{eq:Tgrav}
\end{equation}
with $u^\mu=(1,0,0,0)$. In cylindrical coordinates, the two-dimensional delta function takes the form
\begin{equation}
\delta^{(2)}(\mathbf{x}_\perp)
=
\frac{\delta(r)}{2\pi r}.
\end{equation}
From this perspective, the black string can be naturally interpreted as the gravitational field generated by an infinite line of mass embedded in a cylindrical AdS background. On the other hand, the gauge potential associated with the single copy exhibits singular behavior on the axis $(A_\tau^a \sim 1/r \quad \text{and} \quad F_{\tau r}^a \sim 1/r^2)$, indicating that Maxwell's equations should be interpreted in the distributional sense and written in the presence of sources,
\begin{equation}\label{eq-Maxwell}
\nabla_\mu F^{\mu\nu \, a}
=
J^{\nu \, a}.
\end{equation}
By cylindrical symmetry, the current has only a temporal component. On the gauge-theory side, the color charge per unit length is obtained directly from Gauss's law,
\begin{equation}\label{eq-q}
q^a
=
\int_0^{2\pi}
d\phi\,
\sqrt{-\bar g}\,
F^{r\tau \,a}
=
2\pi c^a\lambda.
\end{equation}
Thus, the current associated with the single copy can be written as
\begin{equation}\label{eq:Jgauge}
J^{\tau \, a}
=
q^a\delta^{(2)}(\mathbf{x}_\perp),
\qquad
J^{r\,a}=J^{\phi \,a}=J^{z\,a}=0.
\end{equation}
This is the most general distributional form compatible with cylindrical symmetry and with a color charge localized on the axis, so that
\begin{equation}
J^{\tau \, a}
=
c^a\lambda\frac{\delta(r)}{r}.
\end{equation}
Finally, we can now explicitly verify that the gravitational source and the gauge current obtained in Eq.~\eqref{eq:Jgauge} satisfy the standard structure of the classical double copy discussed in \cite{Monteiro:2014cda,Carrillo-Gonzalez:2017iyj}. Indeed, the gravitational source given by Eq.~\eqref{eq:Tgrav} is mapped to the single copy current according to
\begin{equation}
J^{\mu a}\propto c^a\lambda u^\mu\delta^{(2)}(\mathbf{x}_\perp),
\end{equation}
such that one of the kinematic factors appearing in the gravitational energy--momentum tensor is replaced by a constant vector in color space,
$$
u_\nu \longrightarrow c^a,
$$
reproducing the general structure of the classical double copy.

Therefore, the correspondence between the gravitational and gauge sources is verified explicitly, in agreement with the general construction of the classical double copy presented in Ref.~\cite{Carrillo-Gonzalez:2017iyj}.

\subsection{The zeroth copy}
The zeroth copy is obtained by removing the KS null vector from the single copy, leaving only the scalar profile $\mathcal{H}(r)$ which characterizes the KS perturbation of the metric. In our case, this scalar function was identified in Eq.~\eqref{eqfs}.  
In the context of the classical double copy in maximally symmetric curved spacetimes, such as the Schwarzschild--(A)dS black hole in $d=4$ spacetime dimensions, the scalar function does not simply satisfy a Laplace equation~\cite{Carrillo-Gonzalez:2017iyj}. Instead, it obeys the differential equation associated with a conformally coupled scalar field \cite{Carrillo-Gonzalez:2017iyj, Luna:2015paa}, which can be written as
\begin{equation}
\left(\bar\nabla^2 - \frac{\bar{R}}{6}\right)\,\phi = j,
\end{equation}
where $\bar{R}$ is the Ricci scalar of the background spacetime. Here, $\phi$ plays the role of our scalar function, while $j$ represents a scalar source. In the study of the black string on a cylindrical AdS background, there is no localized scalar source, and we therefore set $j=0$. Consequently, one expects that in this background the function $\mathcal{H}(r)$ satisfies the equation of a conformally coupled scalar field,
\begin{equation}
\left(\bar\nabla^2 - \frac{\bar{R}}{6}\right)\mathcal{H} = 0.
\end{equation}
Since in this spacetime one has $\bar{R} = 4\Lambda$ (in $d=4$, with $\Lambda<0$), the above equation can be written as
\begin{equation}
\bar\nabla^2 \mathcal{H}(r) - \zeta\, \Lambda\, \mathcal{H}(r) = 0,
\label{biadjoint}
\end{equation}
where $\zeta$ depends on the conventions adopted in the curved-space double copy. For AdS backgrounds, the appropriate choice is $\zeta = 2/3$, in agreement with previous analyses of the double copy in curved spacetime~\cite{Carrillo-Gonzalez:2017iyj}. Computing explicitly the covariant Laplacian on the cylindrical AdS background, and assuming that $\mathcal{H}=\mathcal{H}(r)$, we obtain
\begin{equation}
\bar\nabla^2 \mathcal{H}
=
\frac{1}{\alpha r^2}
\partial_r\!\left( \alpha r^2 \bar g^{rr} \partial_r \mathcal{H} \right)
= -\frac{\lambda \alpha}{ r}.
\end{equation}
Using the fact that in $\mathrm{AdS}_4$ one has $\Lambda = -3\alpha^2$, we verify that the differential equation \eqref{biadjoint} is indeed satisfied, without the need to introduce localized source terms. Thus, in the case of the AdS black string, the curvature of spacetime itself effectively plays the role of a source for the scalar function. This completes the gravity--gauge--scalar correspondence within the framework of the classical double copy, showing that the same scalar function governing the gravitational perturbation in the KS decomposition also consistently determines the structures associated with the single and zeroth copies for the black string.

\subsection{Double copy of the charged black string}

In contrast to the neutral case, the charged black string is a non-vacuum solution of Einstein’s equations, due to the explicit contribution of the electromagnetic energy-momentum tensor. Therefore, once we have established the KS–AdS formulation and the corresponding simple, zero-order copies for the neutral case, we can then extend the same classical double-copy construction to the case of the loaded black string. That is, we consider the static cylindrically symmetric Einstein--Maxwell solution with a negative cosmological constant, starting directly from the result given in Eq.~\eqref{eqF2}. We note that the KS decomposition is completely analogous to the neutral black string case, allowing us to avoid unnecessary repetitions and to make use of the results obtained previously, performing only an extension of the neutral configuration by adding the charge correction term. In \cite{Carrillo-Gonzalez:2017iyj, Cho:2019ype}, a brief construction for black strings and black branes in an $\mathrm{AdS}_d$ background with $d>4$ is presented, starting from a base metric following the formulation of \cite{Chamblin:1999by, Gregory:2000gf, Hirayama:2001bi} for black strings in AdS space. In our case, we adopt the black string approach of \cite{Lemos:1994xp, Lemos:1995cm, Cardoso:2001bb} in a four-dimensional spacetime.

Recently, it has been shown that the interpretation of the classical double copy in the presence of sources requires a more general description than the simple rule valid for dust-like solutions. In particular, contributions to the energy--momentum tensor associated with the energy of fields are not mapped through the usual kinematic replacement but must instead be interpreted through the appropriate contraction of the trace-reversed energy--momentum tensor with a Killing vector of the solution, as discussed in Ref.~\cite{Carrillo-Gonzalez:2017iyj}. In what follows, we show that the charged black string naturally fits into this general framework.

Having established this framework, the construction of the single copy follows directly from the KS decomposition given in Eq.~\eqref{regraSC}, from which we obtain
\begin{equation} \label{eq_campoA_bs_carregada}
A_{\tau}^{a}
=
c^{a}\left(
\frac{\lambda}{\alpha r}
-
\frac{\beta ^2}{\alpha^2 r^2}
\right),
\qquad
A_{r}^{\,a}
=
A_{\phi}^{\,a}
=
A_{z}^{\,a}
=
0.
\end{equation}
As in the neutral case, this expression follows directly from the coordinate representation of the null KS vector in EF coordinates. The corresponding field-strength tensor has only one nonvanishing component, given by
\begin{equation} \label{eq_campoF_bs_carregada}
F_{\tau r}^{a}
=
-\partial_{r} A_{\tau}^{a}
=
c^{a}\left(
\frac{\lambda}{\alpha r^2}
-
\frac{2\beta ^2}{\alpha^2 r^3}
\right).
\end{equation}
The first term corresponds to the contribution associated with the linear mass density of the solution, whereas the second represents the contribution arising from the electric charge, in complete analogy with the Reissner--Nordström--AdS solution discussed in \cite{Carrillo-Gonzalez:2017iyj, Alkac:2021bav}.

To verify the consistency of the single copy obtained above, we first analyze Maxwell's equations on the cylindrical AdS background. Unlike the neutral case, for which the solution satisfies the homogeneous equations throughout the region $r\neq0$, the presence of the electromagnetic field in the gravitational solution implies the existence of a distributed bulk current, as indicated by Eq.~\eqref{eq-Maxwell}. Since the only relevant component is $\nu=\tau$, we can determine the effective current as follows,
\begin{equation}
\nabla_\mu F^{\mu\tau\, a}
=
\frac1{\alpha r^2}
\partial_r
\left(
\alpha r^2F^{r\tau \,a}
\right)
=
J^{\tau \,a}.
\end{equation}
From Eq.~\eqref{eq_campoF_bs_carregada}, it follows that
\begin{equation}
\alpha r^{2}F^{r\tau\,a}
=
\alpha r^{2}\,
c^{a}\left(
\frac{\lambda}{\alpha r^{2}}
-
\frac{2\beta^{2}}{\alpha^{2}r^{3}}
\right)
=
c^{a}\left(
\lambda
-
\frac{2\beta^{2}}{\alpha r}
\right).
\end{equation}
Thus, for all $r\neq0$, we have
\begin{equation}
J^{\tau\,a}_{\rm bulk}(r)
=
\nabla_{\mu}F^{\mu\tau\,a}
=
\frac{2c^{a}\beta^{2}}{\alpha^{2}r^{4}},
\label{eq:JbulkFinal}
\end{equation}
while all remaining components of the current vanish. This distributed contribution is present for all $r\neq0$ and represents, on the gauge-theory side, the contribution induced by the electromagnetic field supporting the charged gravitational solution. However, the gauge potential still contains the term proportional to $\lambda/r$, inherited directly from the neutral black string. As discussed previously, this term exhibits singular behavior on the axis and must be interpreted in the distributional sense, remaining a localized contribution corresponding to an infinite line of color charge. Owing to the linearity of Maxwell's equations, the total current can be decomposed as
\begin{equation}
J^{\tau a}
=
J_{(\lambda)}^{\tau a}
+
J_{(\beta)}^{\tau a}.
\end{equation}
The localized contribution coincides exactly with that obtained for the neutral case, given by Eqs.~\eqref{eq-q} and \eqref{eq:Jgauge}, whereas the distributed contribution is
\begin{equation}\label{eq-contri-eletri}
J_{(\beta)}^{\tau a}
=
\frac{2c^a\beta^2}{\alpha^2r^4}.
\end{equation}
Thus,
\begin{equation}
J^{\tau a}
=
q^a\delta^{(2)}(\mathbf{x}_\perp)
+
\frac{2c^a\beta^2}{\alpha^2r^4},
\end{equation}
or, equivalently,
\begin{equation}
J^{\tau a}
=
c^a\lambda
\frac{\delta(r)}{r}
+
\frac{2c^a\beta^2}{\alpha^2r^4},
\qquad
J^{ra}
=
J^{\phi a}
=
J^{za}
=
0.
\end{equation}

Regarding the interpretation of the sources, unlike the neutral case, the charged solution contains two distinct contributions to the gravitational energy--momentum tensor. Schematically, we may write
\begin{equation}
T_{\mu\nu}^{\rm total}
=
T_{\mu\nu}^{(\lambda)}
+
T_{\mu\nu}^{\rm EM},
\label{eq:Ttotal}
\end{equation}
where $T_{\mu\nu}^{(\lambda)}$ represents the localized source associated with the linear mass density of the black string, whereas $T_{\mu\nu}^{\rm EM}$ corresponds to the energy--momentum tensor of the electromagnetic field distributed throughout the bulk.

Following the general construction for stationary solutions with sources developed in Ref.~\cite{Carrillo-Gonzalez:2017iyj}, the single copy current is obtained from the contraction of the trace-reversed energy--momentum tensor with the Killing vector of the solution. In our case,
\begin{equation}
J^\mu
=
-\frac{2V^\nu}{V^\rho k_\rho}
\left(
T^\mu{}_\nu
-
\frac{\delta^\mu_\nu T}{d-2}
\right),
\label{eq:CarrilloSource}
\end{equation}
where $V^\mu$ is the Killing vector of the spacetime and $T=T^\mu{}_\mu$. Since the solution is static, we choose the timelike Killing vector $V^\mu=\delta^\mu_\tau$. Because the above definition is linear in $T^\mu{}_\nu$, the decomposition \eqref{eq:Ttotal} directly implies the corresponding decomposition of the single copy current,
\begin{equation}
J^{\mu a}
=
J^{\mu a}_{(\lambda)}
+
J^{\mu a}_{(\beta)}.
\label{eq:Jsplit}
\end{equation}
The electromagnetic contribution $T_{\mu\nu}^{\rm EM}$ does not possess the simple structure of a dust-like source, since it depends quadratically on the electromagnetic field tensor. In this case, following the same reasoning adopted for the neutral case, one obtains a distributed bulk current, already determined directly from Maxwell's equations in Eq.~\eqref{eq:JbulkFinal}.

Thus, the decomposition of the gravitational sources given in Eq.~\eqref{eq:Ttotal} is reflected, on the gauge-theory side, by the decomposition of the single copy current expressed in Eq.~\eqref{eq:Jsplit}. In particular, the localized contribution continues to satisfy exactly the standard correspondence of the classical double copy described in Ref.~\cite{Carrillo-Gonzalez:2017iyj}, whereas the distributed contribution represents, on the gauge-theory side, the contribution associated with the electromagnetic energy--momentum tensor present in the gravitational solution. Therefore, the structure of the sources remains consistent with the general construction of the classical double copy: localized sources on the gravitational side give rise to localized sources in the gauge theory, whereas distributed contributions to the energy--momentum tensor are reflected in distributed currents in the single copy.

Finally, the zeroth copy is obtained by removing the null KS vector from the single copy construction, retaining only the scalar function $\mathcal{H}(r)$. In this way, we obtain
\begin{equation}
\mathcal{H}(r)
=
\frac{\lambda}{2\alpha r}
-
\frac{\beta^2}{2\alpha^2r^2}.
\end{equation}
As in the neutral case, this scalar profile corresponds to the zeroth copy of the charged gravitational solution. It is worth noting that the presence of the electric charge modifies only the scalar function through the term proportional to $\beta^2$, while leaving the structure of the KS decomposition, and consequently that of the classical double copy construction, unchanged. As in the neutral case, this profile satisfies the equation of motion of a conformally coupled scalar field on the AdS background. Therefore, the zeroth copy preserves the same structure as the neutral solution, being modified only by the inclusion of the term associated with the electric charge.

\section{CONCLUSION}\label{sec_IV}

In this work, we explore the classical double copy in the context of gravitational solutions with cylindrical symmetry, taking as a case study the static black string of Lemos \cite{Lemos:1994xp, Lemos:1995cm} in an AdS background. Unlike the cases most commonly studied in the literature, which are dominated by spherically symmetric solutions and pointlike sources, black strings constitute gravitational objects with extended horizons and linearly distributed sources. We begin by analyzing the geometric structure of this solution, which can be explicitly written in KS form over an AdS background. This decomposition allows for a clear identification of the scalar profile responsible for the gravitational perturbation, providing the necessary framework for applying the classical double copy formalism. Based on this structure, we construct the associated single copy, obtaining a gauge potential that describes a cylindrical electrostatic field, radial in the transverse plane and uniform along the axial direction of the string.

An analysis of Maxwell’s equations shows that, in the neutral case, the solution is source-free in the bulk for $r \neq 0$, while the $1/r$-type singularity in the gauge potential indicates, in the distributional sense, the presence of a current localized along the axis of the black string, interpreted as an infinite line of color charge, in analogy with the spherically symmetric Schwarzschild case \cite{Monteiro:2014cda}. In turn, the zeroth copy is obtained by removing the KS null vector, and the resulting scalar profile satisfies, on an AdS background, the equation of a conformally coupled scalar field, without the need to introduce additional localized scalar sources, since the spacetime curvature itself effectively closes the equations.

We extend the analysis to the charged black string case by considering the static cylindrically symmetric solution of Einstein--Maxwell theory with a negative cosmological constant. We observe that the inclusion of electric charge affects exclusively the scalar profile of the KS ansatz, preserving both the linearity of the decomposition and the essential structure of the classical double copy. In the single copy sector, the electric charge gives rise to an additional contribution to the gauge current, which manifests itself as a distributed term in the bulk, in addition to the localized source along the axis associated with the mass parameter. In the zeroth copy, we verify that the modified scalar profile continues to satisfy the equation of a scalar field conformally coupled to the AdS background, with no need to introduce additional scalar sources.

As future perspectives, we can naturally highlight an extension of this study to the case of rotating black strings, which can be presented analogously to the spherically symmetric case expressed in Kerr--Newman geometry in standard Boyer--Lindquist coordinates \cite{Lemos:1995cm}, in which angular momentum and the presence of cross-coordinate terms introduce new structures into the KS ansatz, as well as into the interpretation of the single and zeroth copies. Moreover, a particularly interesting scenario that has not been widely explored in the literature concerns regular black strings, especially those arising from black-bounce–type regularizations \cite{Lobo:2020ffi, Furtado:2022tnb, Lima:2022pvc, Bronnikov:2023aya, Lima:2023arg, Lima:2023jtl}, which may provide new avenues for investigating the classical double copy formalism in the presence of smoothed sources. In an analogous manner, these considerations also apply to other regularizations of black holes \cite{Easson:2020esh}, such as those emerging from string T-duality \cite{Nicolini:2019irw, Gaete:2022ukm,Jusufi:2023pzt, Jusufi:2022rbt}, opening up new possibilities for studying the double copy in gravitational geometries  more realistic. As well as several other important applications that have been investigated in the literature within the context of the Kerr--Schild classical double copy.

{\acknowledgments}

The authors would like to thank Conselho Nacional de Desenvolvimento Científico (CNPq) and Fundação Cearence de Apoio ao Desenvolvimento Científico e Tecnológico (FUNCAP) for financial support.


\bibliography{refs}



\end{document}